# Sustainable, low-cost sorbents based on calcium chloride-loaded polyacrylamide hydrogels

Levi Hoogendoorn[1,2*], Mauricio Huertas[1,3*], Phillip Nitz[1], Naiyu Qi[1], Johannes Baller[1], Carsten Prinz[4], Gustav Graeber[1]

*Equal contribution

+gustav.graeber@hu-berlin.de

[1]Graeber Lab for Energy Research, Chemistry Department, Humboldt-Universität zu Berlin, Germany
[2]Integrated Science Program, Materials Science and Engineering, Northwestern University, Evanston, USA
[3]Institute of Chemistry, Chair of Analytical Chemistry, University of Tartu, Tartu, Estonia
[4]Federal Institute for Materials Research and Testing (BAM), Berlin, Germany

**Abstract**

Sorbents are promising materials for applications in atmospheric water harvesting, thermal energy storage, and passive cooling, thereby addressing central challenges related to water scarcity and the global energy transition. Recently, hygroscopic hydrogel composites have emerged as high-performance sorbents. However, many of these systems are fabricated with unsustainable and costly sorbent materials, which hinders their wide deployment. Here, the synthesis of high-performance, cost-efficient polyacrylamide hydrogels loaded with unprecedented amounts of calcium chloride is demonstrated. To this end, the swelling procedure of polyacrylamide hydrogels in aqueous calcium chloride solutions is optimized. The achievable salt loading in the hydrogel is characterized as a function of temperature, calcium chloride concentration in the swelling solution, and the hydrogel preparation conditions. The obtained hydrogel-salt composites are shown to be stable under repeated sorption-desorption cycling and enable water uptakes of 0.92 and 2.38 grams of water per gram of dry materials at 30% and 70% relative humidity, respectively. The resulting cost-performance ratio substantially exceeds lithium chloride-based systems. Further, the mechanistic insights on hydrogel salt interactions can guide the design of sustainable and low-cost sorbent materials for future applications in water and energy.

# 1. Introduction

As our society transitions towards a more sustainable economy, there is a critical need for energy conversion technologies and water treatment solutions that are both economical and environmentally friendly. Functional materials are at the heart of such technologies. To be a good candidate, these materials must be low-cost, eco-friendly, and abundant. However, many crucial components in the contemporary landscape of energy transition technologies depend on raw materials that fail to meet these criteria, such as lithium and cobalt in batteries for electric vehicles, rare earth metals for wind turbine generators, and copper and silver for photovoltaics.[1] Defined as "critical materials for energy" by the U.S. Department of Energy (DOE), these materials require proper substitutions and efficient utilization of new alternatives.

A technology that is particularly important in the global energy transition is sorption, especially in the context of water and heat. With an associated change in enthalpy during the absorption or adsorption of liquid, sorbents are used in various kinds applications such as long-term thermal energy storage,[2] passive cooling,[3] space heating and cooling,[4,5] and atmospheric water harvesting.[6–8] To be attractive for these applications, sorbents should not only be made from sustainable raw materials, but also possess a high water uptake, fast sorption kinetics, and stable cycling capabilities. Instead of the conventional hygroscopic salts such as lithium chloride (LiCl),[9,10] metal-organic frameworks (MOFs),[5,10] zeolites,[11] and silicas,[12] we investigate the application of calcium chloride ($CaCl_2$) in a sorbent system. $CaCl_2$ enables a promising high-performance alternative for sorbents via its widespread availability and excellent hygroscopicity, while bypassing the reliance on lithium and costly fabrication required for MOFs and other sorbents.

Currently, there are a few sorbent systems that incorporate $CaCl_2$, which include $CaCl_2$ in a sodium alginate hydrogel,[13] $CaCl_2$ in a sodium alginate hydrogel with carbon nanotubes (CNTs),[6] and $CaCl_2$ in a polyacrylamide (PAM) hydrogel with CNTs.[14] As the key parameters for sorbents in thermo-adsorptive applications, the water uptake and enthalpy change are proportional to the amount of salt in the sorbent system.[9] Therefore, it is critical to optimize the amount of calcium chloride salt that can be loaded into the hydrogels, especially compared to these previous works. In our work, to reach a maximized salt loading in the system, the PAM hydrogels were first swollen in aqueous $CaCl_2$ solutions at various salt concentrations for 30 days. We obtained key mechanistic insights into the interactions between hydrogels and $CaCl_2$ solutions. This enabled us to optimize the salt loading of the hydrogels as a function of the swelling solution concentration, the swelling time, and the synthesis details of the PAM hydrogels, in order to load unprecedented amounts of salt of more than 17 grams of $CaCl_2$ per gram of PAM into the hydrogels. Such a result exceeds previous loadings by a factor of four. The sorbent system prepared with our optimal swelling procedure shows water uptakes of 0.92, 1.66, and 2.38 grams of water per gram of dry sorbent material at relative humidities (RH) of 30%, 50%, and 70%, respectively. It is noteworthy that the water uptake at 70% RH represents 95% of the water uptake capacity of pure $CaCl_2$. However, in contrast to the use of pure salt, sorbent leakage during cyclic water sorption and desorption can be minimized by incorporating the salt in a hydrogel. Finally, the loading and performance of the $CaCl_2$ PAM system was evaluated at an elevated temperature of 60 °C, where a process of syneresis was observed due to the accelerated hydrolysis of the PAM and therefore the over-crosslinking of the gel when in contact with $Ca^{2+}$ cations in solution. This work shows significant improvements in $CaCl_2$ salt loading in hydrogels as compared to previous studies and reveals new mechanistic insight on adjusting the salt loading. Thereby, our hydrogels show a ratio in water uptake per material cost that substantially exceeds any lithium chloride-based sorbents, making it a promising candidate for wide deployment for energy conversion technologies and water treatment solutions.

## 2. Results and Discussion

### 2.1 Swelling dynamics at room temperature

In **Figure 1**, we present the swelling dynamics of PAM hydrogels in various $CaCl_2$ solutions at room temperature (23 °C). **Figure 1a** shows the swelling ratio as a function of time. The swelling ratio is defined as the current weight of the hydrogel divided by its initial dry weight. A hydrogel in its initial, fully dried state is shown in **Figure 1b**. Starting from this fully dried state, the hydrogels were swollen in five different aqueous $CaCl_2$ solutions. These were 0%$_{sat}$, 25%$_{sat}$, 50%$_{sat}$, 75%$_{sat}$, and 100%$_{sat}$ $CaCl_2$ solutions. Here, %$_{sat}$ refers to the amount of $CaCl_2$ in the aqueous solution, where 0%$_{sat}$ represents pure water, while 100%$_{sat}$ is a fully saturated aqueous $CaCl_2$ solution, based on a reference temperature of 20 °C.[15] The hydrogels reached their equilibrium swelling ratio, defined as the point at which their weight stabilized, at the latest after 30 days of swelling. By swelling the hydrogels in these different solutions, we can reveal the effect of the saturation level on the swelling ratio and thereby on the amount of salt loaded into the hydrogels. Notably, the saturation level of the solution affected both the equilibrium swelling ratio and the kinetics of the swelling process. The equilibrium swelling ratios reached in the 0%$_{sat}$, 25%$_{sat}$, 50%$_{sat}$, 75%$_{sat}$, and 100%$_{sat}$ solutions were 47, 75, 63, 11, and 5, respectively. This reveals a non-linear relationship between the $CaCl_2$ saturation and the equilibrium swelling ratio. The swelling was the highest at low to moderate levels of salt in solution (25%$_{sat}$ and 50%$_{sat}$). Conversely, at high concentrations of salt (75%$_{sat}$ and 100%$_{sat}$), there was only minimal swelling, indicating an unfavorable interaction between the PAM hydrogel and the salt. In the context of loading the hydrogels with as much salt as possible, a higher swelling ratio is favorable. Therefore, swelling in less saturated $CaCl_2$ solutions proves to be conducive towards loading the hydrogels with the maximal amount of $CaCl_2$ sorbent. Besides the equilibrium swelling ratio, the initial kinetics of swelling also differ for swelling in the five solutions. The hydrogels swelling in 0%$_{sat}$ and 25%$_{sat}$ solutions reach 90% of their equilibrium swelling ratio within 2 days of swelling, whereas the hydrogels in the 50%$_{sat}$, 75%$_{sat}$, and 100%$_{sat}$ solutions take 7, 5, and 5 days, respectively, to reach 90% of their equilibrium swelling. Therefore, swelling in more highly saturated $CaCl_2$ solutions slows down the swelling kinetics.

In the remainder of Figure 1, we present images of the hydrogel after swelling in pure water (**Figure 1c**), 25%$_{sat}$ (**Figure 1d**), 50%$_{sat}$ (**Figure 1e**), 75%$_{sat}$ (**Figure 1f**), and 100%$_{sat}$ $CaCl_2$ solutions (**Figure 1g**). The hydrogel in its initial, fully dried state is less than a centimeter in diameter and brittle. After swelling the hydrogel in pure water, it has more than tripled in diameter, is clear in color, and is quite elastic. The hydrogels that were swollen in the 25%$_{sat}$ and 50%$_{sat}$ solutions have also more than tripled in diameter but are slightly cloudy, on account of the salt. Hydrogels exposed to the 75%$_{sat}$ and 100%$_{sat}$ solutions remain small, clear, and brittle, which corresponds with the lack of swelling in these solutions.

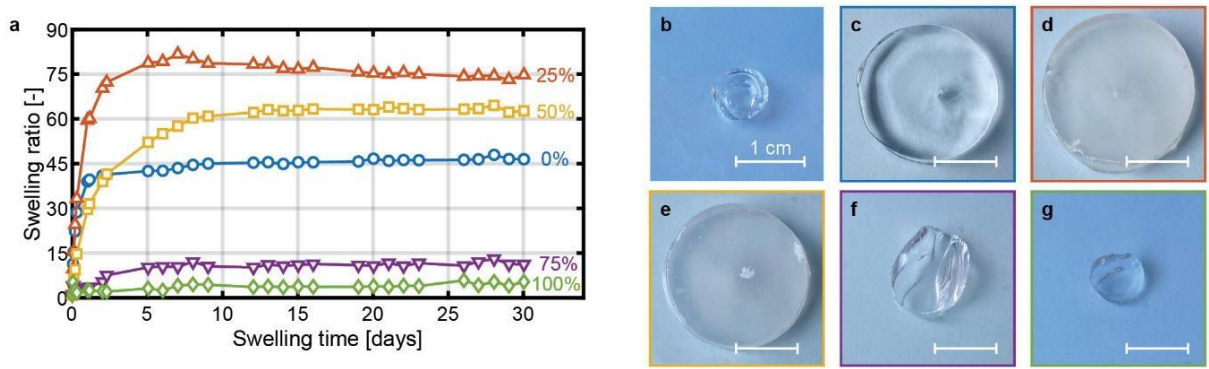

**Figure 1. Swelling dynamics of PAM hydrogels in aqueous CaCl$_2$ solutions at room temperature. a**, Swelling ratio, defined as current sample weight divided by the initial dry sample weight, as a function of swelling time, for samples swollen in aqueous CaCl$_2$ solutions of varying concentration of 0%$_{sat}$, 25%$_{sat}$, 50%$_{sat}$, 75%$_{sat}$, and 100%$_{sat}$, respectively. **b**, Picture of dry hydrogel before swelling. **c-g**, Pictures of the hydrogels after swelling for 30 days in pure water, i.e., 0%$_{sat}$ solution (c), 25%$_{sat}$ solution (d), 50%$_{sat}$ solution (e), 75%$_{sat}$ solution (f), and 100%$_{sat}$ solution (g), respectively.

**2.2 Salt loading at room temperature**

After swelling the hydrogels in the five different CaCl$_2$ solutions, the increased mass of the swollen hydrogels consisted of a combination of both water and salt. In terms of the sorbent performance of the CaCl$_2$ PAM system, it is important to characterize the amount of CaCl$_2$ that was loaded in the hydrogel, see **Figure 2**. The salt loading is typically expressed as the mass of salt, g$_{CaCl2}$, divided by the mass of polymer, g$_{PAM}$. To determine the salt loading, we can assume that the salt concentration of the aqueous solution is the same inside and outside the hydrogel network.[9] In the computation, we first determine the amount of salt solution that was taken in by the gel matrix and normalized it to the initial mass. Then, we multiply this value by the mass fraction of CaCl$_2$ in the solution to determine the salt loading (see Supporting Information for details). These computed values are shown in **Figure 2a**. We find that the hydrogels in the 25%$_{sat}$, 50%$_{sat}$, 75%$_{sat}$, and 100%$_{sat}$ solutions had salt loadings of 12.7, 17.3, 4.3, and 2.1 g$_{CaCl2}$ g$_{PAM}^{-1}$, respectively. We note that the hydrogel with the highest equilibrium swelling ratio (from the 25%$_{sat}$ CaCl$_2$ solution) does not have the highest salt loading, due to the difference in the solution concentration. The SEM image in **Figure 2a** clearly shows the salt on the surface of the hydrogel swollen in the 50%$_{sat}$ CaCl$_2$ solution, after it was fully dehydrated for imaging.

We can compare the salt loading in our hydrogel system to other CaCl$_2$-based sorbent systems in the literature in **Figure 2b**. We loaded the PAM hydrogel with more than 17 g$_{CaCl2}$ g$_{PAM}^{-1}$, demonstrating the highest salt loading to date in the literature, for CaCl$_2$ in hydrogels; we compared this to a salt loading of 4 g$_{CaCl2}$ g$_{Hydrogel}^{-1}$ in a PAM system with CNTs,[14] 3.17 g$_{CaCl2}$ g$_{Hydrogel}^{-1}$ in a sodium alginate hydrogel system,[13] and 0.5 g$_{CaCl2}$ g$_{Hydrogel}^{-1}$ in another sodium alginate system but with CNTs.[6]

We also used thermogravimetric analysis (TGA) to confirm the calculated salt content. In **Figure 2c**, we present the relative mass of pure CaCl$_2$ salt as the temperature increases to 800 °C. From this, we determined that the salt was fully dry at 235 °C. This temperature closely matches the literature, which found that CaCl$_2$ was fully dehydrated at 240 °C under an inert atmosphere.[16] This temperature of 235 °C is used as a reference in the following TGA experiments. In **Figure 2d**, the relative mass of a salt-free PAM hydrogel partially swollen in pure water is shown. In the literature, we find that PAM is fully dehydrated at 220 °C and that it can start to degrade above this temperature with irreversible pyrolysis.[17,18] Finally, in **Figure 2e**, the relative mass of the PAM hydrogel swollen in the 50%$_{sat}$ CaCl$_2$ solution is shown as a function of temperature in the TGA. We determine the salt loading based on the relative mass of the CaCl$_2$/PAM/water TGA curve at 235 °C from **Figure 2e**. At this point, the CaCl$_2$ and the PAM are both fully dry and there may be minor loss of PAM mass on account of the thermal degradation, which leads to a conservative estimate of the salt content. From this data point, we determine the experimental salt loading to be 18.6 g$_{CaCl2}$ g$_{PAM}^{-1}$, which is within a 10% deviation as compared to the computed salt loading. Overall, the TGA confirms that unprecedented levels of CaCl$_2$ salt content can be achieved when swelling the hydrogel in a suitable CaCl$_2$ solution for sufficiently long time to reach equilibrium. These results are therefore promising for use of the CaCl$_2$ PAM system as an effective sorbent.

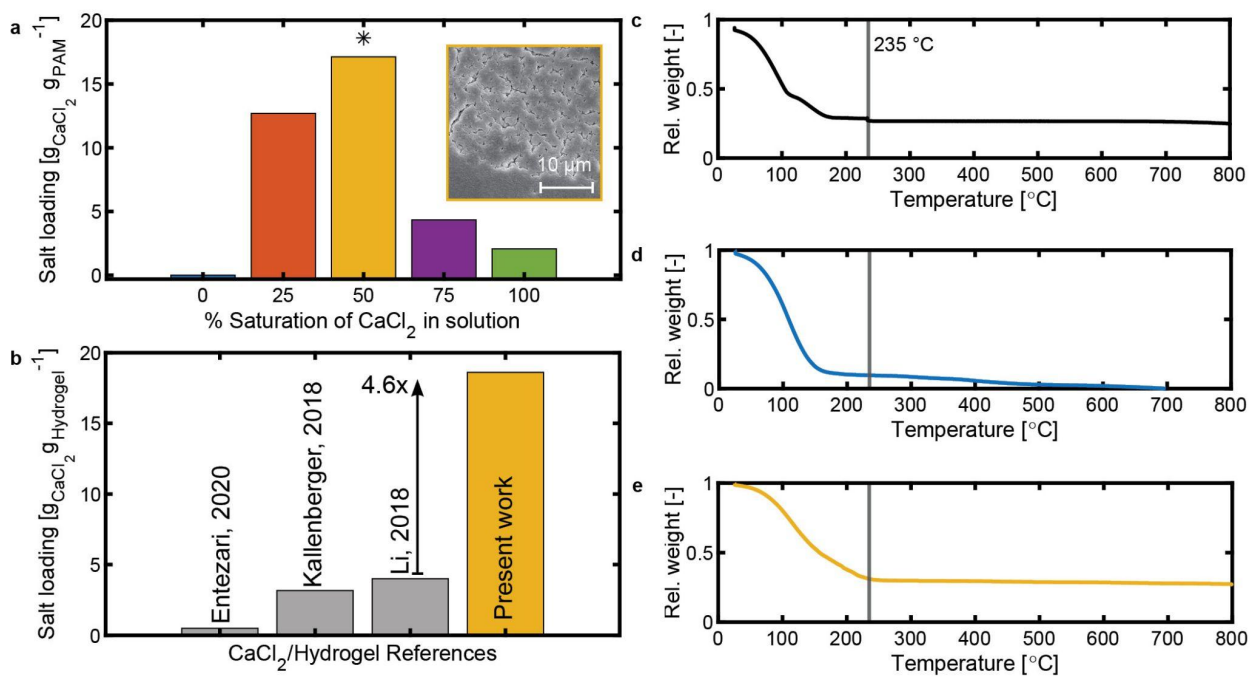

**Figure 2. Salt loading in hydrogels at room temperature. a**, Predicted salt loading of the hydrogels swollen to equilibrium in the various aqueous CaCl$_2$ solutions as a function of the salt content defined in % saturation. The asterisk defines the experimentally determined salt loading of the hydrogel swollen in the 50%$_{sat}$ CaCl$_2$ solution, as measured by thermogravimetric analysis (TGA). The inset shows the surface of the hydrogel swollen in the 50%$_{sat}$ CaCl$_2$ solution, with salt crystals apparent on the surface. **b**, Comparison of the experimentally-validated salt loading of 18.6 g CaCl$_2$ per g PAM (as compared to the calculated salt loading of 17.6 g CaCl$_2$ per g PAM as shown in panel a) from the present work with previous CaCl$_2$ hydrogel systems, including systems presented by Entezari et al. (sodium alginate with CNT hydrogel),[6] Kallenberger et al. (sodium alginate hydrogel),[13] and Li et al. (polyacrylamide with CNT).[14] **c**, Relative weight, defined as current sample weight divided by initial sample weight at the beginning of the analysis, of an aqueous 50%$_{sat}$ CaCl$_2$ solution as a function of temperature obtained via TGA. **d**, Relative weight versus temperature of a salt-free polyacrylamide hydrogel partially swollen in water. **e**, Relative weight of a hydrogel swollen to equilibrium in 50%$_{sat}$ CaCl$_2$ solution versus temperature.

## 2.3 Mechanistic insights into swelling at room temperature

From the previous analysis, we see that both the kinetics and equilibrium swelling ratio depend on the concentration of the CaCl$_2$ solution. In **Figure 3**, we show several key experiments that shed light on the swelling process at room temperature. First, in **Figure 3a**, we present the elastic modulus, $E$, of a PAM hydrogel as a function of the swelling ratio in pure water (i.e., 0%$_{sat}$ CaCl$_2$). $E$ was determined from flat punch indentation testing. From the data, we can establish that $E$ generally decreases as the swelling ratio increases. At a swelling ratio of 12, $E$ is 18.9 kPa, and at a swelling ratio of 59, $E$ decreases to 4.9 kPa. Importantly, $E$ can be used to characterize the properties of the polymer network. Specifically, $E$ is correlated to the number of moles of elastic chains per unit volume of the network.[19,20] As a result, $E$ can approximate the network density at various swelling ratios. At low swelling ratios, $E$ is high and therefore the density of chains is also high. As the hydrogel swells, $E$ decreases, the volume of the hydrogel increases, and therefore the density of chains decreases. Qualitatively, the mesh size, defined as the average distance between two neighboring network junctions, and therefore the permeability of the PAM hydrogel, decreases as the effective crosslinking is increased in a more dense network.[21] In terms of loading the hydrogels, such a connection is important, as we can better understand how the polymer network behaves during swelling. In part, the polymer network density dictates the mesh size within the polymer network. The relation between the equilibrium mesh size between polymer chains, $\xi_\infty$, and $E$ is given by equation (1):[9]

$$\xi_\infty = \left(\frac{2kT(\nu + 1)}{E}\right)^{1/3} \qquad (1)$$

where $k$ is Boltzmann's constant, and $T$ is the temperature (here ~298 K), and $\nu$ is Poisson's ratio (here assumed to be 0.5). At the upper and lower bounds of our experimentally determined elastic moduli of 18.9 kPa and 4.9 kPa, we therefore obtain an equilibrium mesh size between 8.7 nm and 29.3 nm. As discussed below, this is especially important for hydrogels swelling in highly concentrated aqueous CaCl$_2$ solutions.

Aside from the mechanical testing, we also performed hydrogel swelling experiments in additional CaCl$_2$ solutions. We swelled hydrogels in 55%$_{sat}$, 60%$_{sat}$, 65%$_{sat}$, and 70%$_{sat}$ CaCl$_2$ solutions, in addition to the five solutions tested in **Figure 1**. In **Figure 3b**, the equilibrium swelling ratios (after 30 days of swelling) are shown. After swelling in 55%$_{sat}$, 60%$_{sat}$, 65%$_{sat}$, and 70%$_{sat}$ solutions, the hydrogels reached equilibrium swelling ratios of 36, 20, 19, and 18, respectively. The equilibrium swelling ratio for low saturation solutions (25%) is therefore much higher than swelling in more concentration solutions. Specifically, we confirm that swelling in solutions with a CaCl$_2$ saturation higher than 60%$_{sat}$ is very limited.

In **Figure 3c**, we swelled PAM hydrogels in the same five solutions as in **Figure 1**, but we changed the initial condition. In **Figure 1**, the hydrogels were synthesized, cured for more than 24 hours, and then fully dried in an oven before starting swelling. In contrast, for the data shown in **Figure 3c**, the hydrogels are synthesized, cured, and then used directly in the swelling experiment without the drying step. These as-cured hydrogels already contain some amount of water from the synthesis process. Thus, the as-cured hydrogels start from a swelling ratio of approximately 13, even before swelling in the CaCl$_2$ solutions. By changing the initial condition, we find that the equilibrium swelling ratio can be drastically increased. The comparison of swelling from a fully dry state versus from an as-cured state is presented in **Figure 3c**. After swelling from the as-cured state, the hydrogels in the 0%$_{sat}$, 25%$_{sat}$, 50%$_{sat}$, 75%$_{sat}$, and 100%$_{sat}$ solutions reach equilibrium swelling ratios of 50, 71, 71, 46, and 28. Compared to the hydrogels swollen from a fully dry state, we see a change of +6%, -5%, +23%, +318%, and +460% in the swelling ratio. Clearly, the difference in the equilibrium swelling ratio by altering the initial condition is much more pronounced in the higher saturated CaCl$_2$ solutions. The small changes in swelling ratio at lower saturated solutions can likely be attributed to some variance in the hydrogel swelling per sample. These experiments therefore provide important insights into how the polymer network and how the concentration of the CaCl$_2$ solution can affect the swelling of the hydrogel.

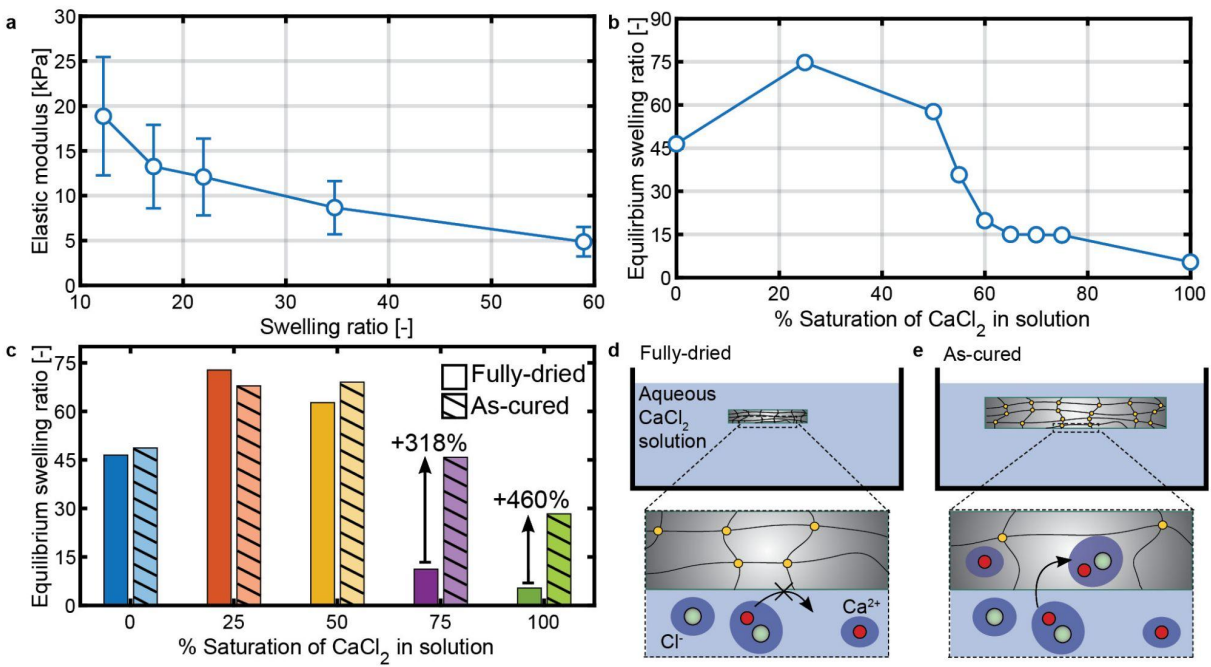

**Figure 3. Mechanistic insights into the swelling of PAM hydrogels in aqueous CaCl$_2$ solutions at room temperature. a**, Elastic modulus, $E$, of a salt-free PAM hydrogel as a function of its swelling ratio. $E$ can be correlated to the concentration of elastically effective chains, and therefore be used to determine the network density at various swelling ratios. **b**, Equilibrium swelling ratio (after 30 days) of initially fully-dried hydrogels in aqueous CaCl$_2$ solutions as a function of salt concentration in the solution. **c**, Equilibrium swelling ratio after 30 days of swelling as a function of the initial condition of the hydrogel: initially fully-dried (plain bar) and as-cured hydrogels (hatched bar) in 0%$_{sat}$, 25%$_{sat}$, 50%$_{sat}$, 75%$_{sat}$ and 100%$_{sat}$ CaCl$_2$ solutions. Starting the swelling process from the as-cured state enables substantial enhancements in the equilibrium swelling ratio in highly concentrated solutions. **d**, Schematic showing the swelling mechanism for a collapsed, fully-dried hydrogel in a highly saturated CaCl$_2$ solution. The solvated calcium and chloride ions cannot enter the hydrogel effectively due to larger solvated complexes and ion pairing at higher concentrations. **e**, Schematic showing the swelling mechanism for a more open as-cured hydrogel in a highly saturated CaCl$_2$ solution. The larger solvated ions can enter the more open and less dense hydrogel structure, therefore swelling the hydrogel.

In **Figure 3d-e**, we present a mechanism that ties together the insights from these additional experiments. In **Figure 3d**, we show the swelling mechanism of an initially dry hydrogel, in a highly saturated CaCl$_2$ solution. In such a solution, the calcium (Ca$^{2+}$) and chloride (Cl$^-$) ions are fully solvated and there is a lack of free water.[22] In addition, the calcium and chloride ions can form ion pairs and therefore agglomerate to larger solvated complexes.[23] Such a phenomenon occurs in highly saturated solutions on account of the density of ions within the solutions. We propose that the solvated complexes therefore cannot enter the dry hydrogel, which has a high density of elastic polymer chains and consequently a lack of open space and a small mesh size for the ions to enter.

Evidence for this proposed mechanism can be obtained by comparing the size of the ions with the pore size of the hydrogel. The size of the first hydration shell of solvated $Ca^{2+}$ ions in water is around 0.25 nm, with 7 to 8 coordinated water molecules.[23–26] The second hydration shell has a radius of 0.47 nm, with up to 20 coordinated water molecules.[23,24] In comparison, the hydration shell of $Cl^-$ ions in an aqueous solution is between 0.314 nm and 0.398 nm.[23,27] Further, at higher concentrations of $CaCl_2$ in solution (i.e., at 4.0 and 6.4 M in the literature), paired ions can lead to larger complexes of solvated ions.[23] Relative to our concentrations of salt, note that 50%$_{sat}$ $CaCl_2$ is equivalent to 3.26 M, 75%$_{sat}$ is equivalent to 5 M, and 100%$_{sat}$ is equivalent to 6.7 M. Therefore, these values suggest that the solvated complexes that attempt to enter the hydrogels at these higher concentrations are physically larger and may therefore not diffuse as readily into the hydrogel. For example, for a conservative estimate we assume that the first solvated shells of the $Ca^{2+}$ and $Cl^-$ are just overlapping in an ion pairing condition. Consequently, we can estimate the total solvated complex to have a radius of approximately 0.8 nm, with up to 40 coordinated water molecules.

For comparison, we can estimate the pore size within the hydrogels. First, we can make an estimate from the equilibrium mesh size computed above with equation (1). We can divide the mesh size values in half to estimate the radius of the pores to be between 4.4 nm and 14.7 nm. Second, we can estimate the radius of the pores starting from the literature. For PAM hydrogels, the reported pore sizes range between 20 and 200 nm, depending on the synthesis and the amount of crosslinker.[28,29] In these references, the hydrogels are often in their as-cured or fully swollen state. In our experiments, when the hydrogel starts from a fully swollen state, the gel can be a factor of around 3 smaller in the as-cured state, and from the as-cured state, it can be another factor of 11 smaller in the fully dry state (see Supporting Information for further details). Therefore, if we consider a range of between 4 and 200 nm pore sizes in the fully swollen state based on our present work and the references, then we would estimate a range of 1.33 nm and 66.67 nm in the as-cured state, and pore sizes between 0.12 nm and 6 nm in the fully dry state. In this calculation, we assume that the hydrogel shrinks isotropically and that the pore size scales with the volume of the hydrogel. We can therefore conclude, that the pore size of a fully dry PAM hydrogel can in fact be of a similar size as the ion pairing radius of 0.8 nm. This supports our theory of how fully dry and as-cured hydrogels swell differently in higher saturated $CaCl_2$ solutions.

In contrast, in **Figure 3e**, we sketch an as-cured hydrogel, which is already partially swollen at the start of the swelling process. Here, ions are at the same concentration and level of solvation, but the hydrogel network density is lower and the average pore size is larger. As a result, the ions can enter the hydrogel and continue swelling the structure. Thus, such a mechanism considers the change in $E$ as a function of the swelling ratio. In addition, the mechanism explains why swelling in highly concentrated solutions, where there are larger solvated ion complexes, is less effective. In less concentrated solutions there is more free water available and less large, solvated complexes. The water could first enter the hydrogel and increase the swelling ratio. At a certain increased swelling ratio, where the polymer network is more open, the ions could then enter the hydrogel. However, in highly concentrated solutions, such free water is not available and there is a size limitation for the ions to enter smaller pores of the hydrogel. The mechanism also accounts for why starting in the as-cured state is especially important for swelling in highly concentrated solutions. In the as-cured state, the polymer network is already more open, and the solvated ions can enter the hydrogel immediately. Overall, it is important to tune both the saturation of the $CaCl_2$ solution and the initial condition of the polymer network when loading polyacrylamide with $CaCl_2$ through a swelling mechanism. For instance, the as-cured hydrogel swollen in the 50%$_{sat}$ CaCl$_2$ solution has a calculated salt loading of 21 g $_{CaCl2}$ g $_{PAM}$$^{-1}$. Thus, the salt loading can be improved by altering these conditions.

**2.4 Performance of CaCl$_2$ PAM system as a sorbent**
In **Figure 4**, we present the performance metrics of the CaCl$_2$ PAM system as a sorbent. We also compare the performance to other systems in the literature. In determining the performance, we consider our highest salt-loaded hydrogel, which was the hydrogel swollen in the 50%$_{sat}$ CaCl$_2$ solution. We first test the water uptake of the salt-loaded hydrogel using dynamic vapor sorption (DVS). In **Figure 4a**, we show the water uptake of the hydrogel, defined as the mass of water divided by the initial dry weight of the sample, at RH between 10% and 90%. A higher water uptake is reached at higher RH, with a water uptake of 3.7 g$_{H2O}$ g$_{dry}$$^{-1}$ at 90% RH. For comparison, we also plot the water uptake that we measure for pure CaCl$_2$. The loaded hydrogel reaches a similar sorption to the pure CaCl$_2$, demonstrating the efficacy of the CaCl$_2$ PAM system as a sorbent. In contrast to the pure salt, however, the CaCl$_2$ PAM system does not suffer from leakage and deliquescence.[9] Instead, the hydrogel matrix can store the captured salty water, making the CaCl$_2$-loaded PAM a powerful engineering material for applications ranging from water harvesting to thermo-adsorptive energy storage, where the pure salt could not be used. The water uptake behavior that we find for our CaCl$_2$ PAM system aligns with other hydrogel systems loaded with sorbents where the hygroscopicity is largely determined by the sorbent in the hydrogel.[9,14]

In **Figure 4b**, we demonstrate our set-up for testing the cycling capabilities of the salt-loaded hydrogel when alternately exposed to a humid and a dry environment, respectively. To create the humid environment, we use a desiccator as shown in **Figure 4b**. The closed desiccator contains a beaker at the bottom that is filled with a fully saturated NaCl solution. The hydrogel sample is placed on a metal grid that is located above the beaker filled with the NaCl solution. There is no contact between the hydrogel and the NaCl solution and the only purpose of the NaCl solution is to maintain a constant atmosphere in the desiccator with a RH of 74% at room temperature. To create the dry environment, we use an oven set to 60 °C. The loaded hydrogel is placed in the oven for 16 hours and then in the desiccator for 8 hours, for 4 cycles, to determine the performance over several cycles. In **Figure 4c**, the water uptake as a function of time is plotted. The hydrogel therefore reaches a water uptake of around 2 g$_{H2O}$ g$_{dry}$$^{-1}$ in each cycle, with no substantial decrease in performance.

Altogether, we have demonstrated an effective and scalable sorption system based on the sustainable salt of CaCl$_2$. To fully understand the added benefit of using CaCl$_2$ for a sorption system, we want to estimate the cost-performance ratio of our material and compare it to our recent lithium chloride-based sorbent.[9] To compute the cost-performance ratio, we define performance as the water uptake at 70% RH in g$_{H2O}$ g$_{dry}$$^{-1}$, and cost based on the material cost of the dry sorbents in USD g$_{dry}$$^{-1}$. For the present work and our previous work, more than 90% of the mass of the hydrogel-salt composite is contributed by the hygroscopic salt, while the remainder is mostly low-cost monomer. As a result, the material cost of these two composites can be estimated by the cost of salt only. Based on recent global market prices, we estimate the material cost of CaCl$_2$ to be around USD 220 per metric ton of dry material, while for LiCl, we determine an average price of around USD 110000 per metric ton of dry material (see Supporting Information for details). Considering the water uptakes of our CaCl$_2$-loaded PAM composite and the previously reported LiCl-loaded PAM composite at 70% RH to be 2.38 g$_{H2O}$ g$_{dry}$$^{-1}$ and 3.86 g$_{H2O}$ g$_{dry}$$^{-1}$, respectively, we determine a cost-performance ratio of 10818 g$_{H2O}$ USD$^{-1}$ for our present work, while it is only 35 g$_{H2O}$ USD$^{-1}$ for the LiCl-loaded PAM. In summary, the cost-performance ratio of our CaCl$_2$-loaded PAM is more than three orders of magnitude higher than a high-performance sorbent using LiCl. This underlines why the use of abundant and low-cost raw materials such as $CaCl_2$ is attractive for the wide deployment of sorbents in various applications ranging from water to energy.

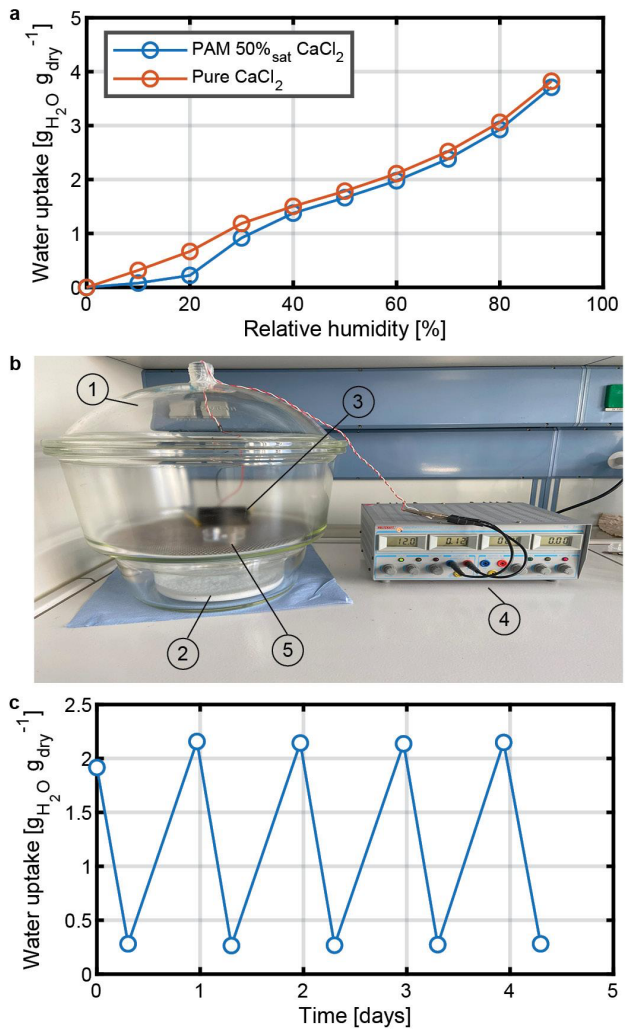

**Figure 4. Water harvesting performance of the $CaCl_2$-loaded PAM hydrogel. a**, Water uptake, defined as mass of water in the material per gram of dry sample including PAM and salt, measured with dynamic vapor sorption device, comparing the uptake of a hydrogel swollen for 30 days in 50%$_{sat}$ $CaCl_2$ solution and pure $CaCl_2$. **b**, Image of desiccator set-up, where (1) points to the desiccator itself, which is a fully closed environment, (2) is the fully saturated NaCl solution inside the desiccator, bringing the environment to a relative humidity of 74%, (3) is the fan inside the desiccator to promote a homogeneous environment, (4) is the power source for the fan, and (5) is a dish containing the $CaCl_2$ loaded hydrogel. **c**, Cyclic water uptake over the course of four days of a hydrogel prepared by swelling it for 30 days in 50%$_{sat}$ $CaCl_2$ solution. The sample is exposed alternately to a humid condition (74% RH, room temperature) for 8 hours, and to a dry condition (60 °C in the oven) for 18 hours.

## 2.5 Swelling dynamics at an elevated temperature

In addition to swelling the hydrogel at room temperature, we also conducted swelling experiments at an elevated temperature of 60 °C. Previous work has shown that increasing the swelling temperature can enhance the swelling kinetics and the equilibrium swelling ratio, which would be important in increasing the salt loading and ultimate performance of the system.[9] For these experiments, fully-dried hydrogels as shown in **Figure 1b**, were placed into the same saturated solutions as those at room temperature. However, for the duration of the swelling, the samples were placed in the oven at 60 °C. The swelling ratio as a function of time is presented in **Figure 5a**. After swelling in 0%$_{sat}$, 25%$_{sat}$, 50%$_{sat}$, 75%$_{sat}$, and 100%$_{sat}$ $CaCl_2$ solutions at 60 °C for 30 days, the hydrogels reached equilibrium swelling ratios of 79, 4, 4, 4, and 6, respectively. While the hydrogel in pure water swelled more at an elevated temperature, the hydrogels in the various $CaCl_2$ solutions reached very small final equilibrium swelling ratios. At elevated temperatures, the initial kinetics of swelling were increased for all samples. For example, the hydrogel in the 50%$_{sat}$ solution reached a swelling ratio of 60 in one day at the elevated temperature, compared to eight days at room temperature. However, after reaching some maximum swelling ratio, all the hydrogels in the $CaCl_2$ solutions started to shrink and collapse. The collapse can be clearly demonstrated with the images of the hydrogels in **Figure 5b-g**. In **Figure 5b**, the hydrogel that was swollen in pure water is shown. The hydrogel is large and clear and looks identical to the hydrogel swollen at room temperature. In **Figure 5c-f**, the hydrogels swollen in 25%$_{sat}$, 50%$_{sat}$, 75%$_{sat}$, and 100%$_{sat}$ solutions at 60 °C are shown after being in the $CaCl_2$ solutions for 30 days. These hydrogels are all small, similar in diameter to the initial dry state, and cloudy. In contrast, when removing the hydrogel after only five days from the $CaCl_2$ solution, the hydrogels do not collapse (**Figure 5g).** The hydrogels that collapsed are undergoing a process of hydrolysis and then syneresis. At room temperature, as shown in **Figure 5h**, the PAM hydrogels consist of polymer chains with amide side groups. However, at elevated temperatures, such as 60 °C, the PAM can hydrolyze to become partially hydrolyzed polyacrylamide (PHPA).[30–32] In this process, some of the amide side chains of the polymer hydrolyze to become anionic carboxylic acid. More than 50% of the polymer chain side groups can be hydrolyzed at this temperature.[31] When the PHPA is in contact with an aqueous calcium chloride solution, there is an attractive interaction between the anionic side groups and the cationic calcium ions in solution within the hydrogel. This interaction causes over-crosslinking of the PHPA whereby the chains are pulled in together towards the calcium ions on account of the electrostatic interaction, in a process termed syneresis, as shown in **Figure 5i**.[30] Therefore, the combination of accelerated hydrolysis at high temperatures and the presence of calcium ions in the polymer structure causes the hydrogel to collapse. In applications such as thermal energy storage, this is an especially important consideration. In energy storage, the $CaCl_2$ PAM sorbent system would be subject to varying temperatures and need to be thermally stable.[33] Therefore, further work could focus on the analysis of $CaCl_2$ PAM sorbents within thermal energy applications, to investigate how the process of syneresis may affect the long-term stability of the material.

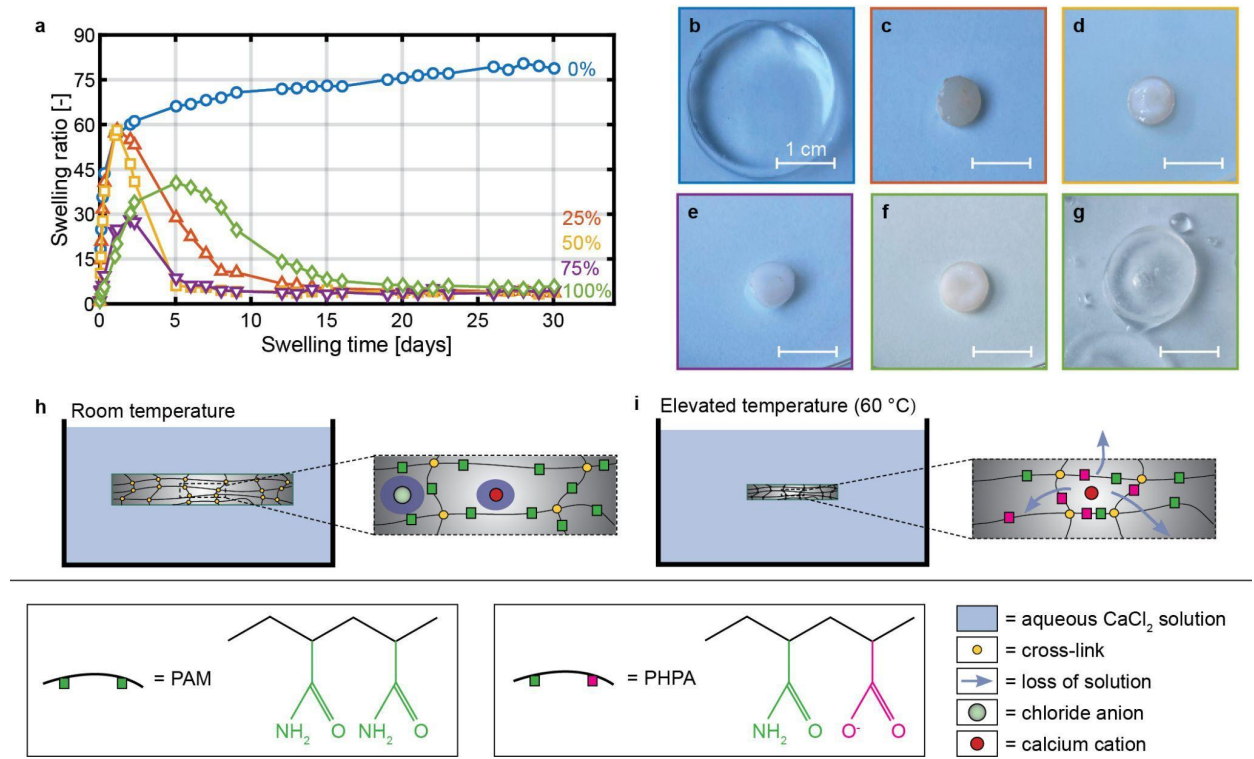

**Figure 5. Swelling dynamics and collapse due to syneresis of PAM hydrogels at an elevated temperature of 60 °C. a**, Swelling ratio as a function of swelling time at 60 °C, for samples swollen in 0%$_{sat}$, 25%$_{sat}$, 50%$_{sat}$, 75%$_{sat}$, and 100%$_{sat}$ CaCl$_2$ solutions. The degree of salt saturation in the solution is computed based on a reference temperature of 20 °C. **b-f**, Pictures of the corresponding hydrogels after swelling for 30 days in pure water, i.e., 0%$_{sat}$ solution (b), 25%$_{sat}$ solution (c), 50%$_{sat}$ solution (d), 75%$_{sat}$ solution (e), and 100%$_{sat}$ solution (f), respectively. **g**, Picture of a hydrogel after swelling for only 5 days in 100%$_{sat}$ solution. **h**, Schematic illustrating hydrogel swelling at room temperature without hydrolysis of the polyacrylamide hydrogel. **i**, Schematic showing how PAM hydrolyzes to partially hydrolyzed polyacrylamide (PHPA) at elevated temperatures. PHPA features side groups that are hydrolyzed to anionic carboxylic acid. These negatively charged side chains can interact with the calcium ions inside the hydrogel, causing over-crosslinking due to the electrostatic interaction, leading to the expulsion of water and collapse of the hydrogel in a process termed syneresis.

## 3. Conclusion

In this work, we performed swelling experiments of PAM hydrogels in aqueous CaCl$_2$ solutions. We varied the salt concentrations in the swelling solutions to determine the optimal synthesis conditions for high-performance hydrogel-salt composites. We achieved an unprecedented amount of CaCl$_2$ salt loading in a hydrogel of more than 17 g CaCl$_2$ per gram of polymer, which is four times more than previous studies had shown. Based on systematic swelling experiments at room temperature, we provided critical mechanistic insights for the interaction between hydrogels, CaCl$_2$, and water, and also demonstrated the importance of the initial conditions of the hydrogel at the start of swelling in determining the final uptake of water and salt. The cyclic stability of the CaCl$_2$ PAM composites was tested by repeated sorption-desorption cycling. The water uptake was measured via dynamic vapor sorption to be 0.92 g$_{H2O}$ g$_{dry}^{-1}$, 1.66 g$_{H2O}$ g$_{dry}^{-1}$ and 2.38 $g_{H2O}$ $g_{dry}^{-1}$, at relative humidities of 30%, 50%, and 70%, respectively. Due to its high hygroscopicity obtained exclusively with abundant, low-cost materials, we found that our sorbent provides an outstanding cost-performance ratio, which is more than three orders of magnitude larger than other synthesized sorbents relying on the less cost-effective lithium chloride. Besides the room temperature experiments, we conducted systematic swelling experiments at an elevated temperature of 60 °C. At a higher temperature, we found that the hydrogels underwent hydrolysis and showed syneresis via an over-crosslinking mechanism when in contact with calcium ions in solution. Overall, this work provides a significant step towards understanding the swelling kinetics and final salt loading of hydrogel sorbents, which is of critical importance to various applications ranging from atmospheric water harvesting to thermal energy storage, and passive cooling.

## 4. Experimental section

**Materials.** Acrylamide (≥ 98.5% pure) and calcium chloride (> 93% pure and anhydrous) were purchased from Thermo Scientific. N,N'-methylenediacrylamide (Bis-acrylamide), ammonium persulfate (APS, ACS reagent, ≥ 98.0% pure) and N,N,N',N'-tetramethylethylenediamine (TEMED) were purchased from Merck/Sigma Aldrich. HPLC water (≤ 1μS cm$^{-1}$) and demineralized water (≤ 1.5 μS cm$^{-1}$) were purchased from VWR chemicals. All chemicals were used without further purification.

**Hydrogel synthesis.** Neutral PAM hydrogels were synthesized by the polymerization of acrylamide monomers with bis-acrylamide crosslinker based on the one-pot synthesis procedure described in Graeber, et al. 2023[9]. Before starting the synthesis, stock solutions of the crosslinker (i.e., bis-acrylamide) and the initiator (i.e., APS) were prepared. For the crosslinker stock solution, 25 mg of bis-acrylamide was mixed with 10 g of HPLC water in a closed glass container. For the initiator stock solution, 71 mg of APS was mixed with 10 g of HPLC water in a closed glass container. To start the synthesis, 8.36 g of acrylamide was mixed in 96 g of HPLC water at room temperature. Since acrylamide is toxic, it is important to perform this step in a fume hood using appropriate personal protective equipment. Subsequently, 2 g of the bis-acrylamide stock solution and 2 g of the APS stock solution were added, while stirring continuously. Finally, 12 μL of TEMED was added to the solution. The liquid pre-gel was poured into 15 mL polypropylene plastic containers (VWR) and sealed for curing for 24 hours. We refer to such a sample, which is obtained after 24 hours of curing, as the as-cured sample. The contact with air was minimized during the curing process by filling the vials to the top, as oxygen can interfere with the polymerization process of the gel. The hydrogel disks used for the subsequent swelling experiments were obtained by cutting the gel with ceramic scissors. After cutting, the disks were stored in demineralized water. For the swelling experiments, disks with similar masses (relative deviation of the group within 3%) were selected. Unless otherwise noted, the selected disks were dried in an oven at 60 °C for at least 3 days. Only for the experiments shown in **Figure 3c**, the discs were used in the as-cured state (without the final drying step).

**Swelling experiments.** $CaCl_2$ solutions were prepared based on percentages of the saturation concentration at a reference temperature of 20 °C for a standard aqueous solution of the salt. At 20 °C, saturation occurs at 72.8 g of $CaCl_2$ per 100 g of water.[15] Solutions of different concentrations were prepared in demineralized water from 25%$_{sat}$ to 100%$_{sat}$. Since the mixing of $CaCl_2$ and water is exothermic, the aqueous solutions heated as the salt is solubilized. The solutions were continuously stirred until they cooled down again to room temperature. For consistency, we also kept the same labels for the high temperature solutions, still referring to their degree of saturation based on the reference temperature of 20 °C. However, the percentage of saturation at a reference temperature of 60 °C is effectively smaller since the salt concentration to reach saturation at 60 °C is 131.1 g $CaCl_2$ per 100 g of water.[15] To swell the gels, ~30 mL of the CaCl$_2$ solutions were poured into 50 mL polypropylene plastic containers and the dried gels were added to the solutions. Over the course of the swelling experiments, the gels were briefly removed from the swelling solution each day and weighed on a scale to monitor the sample weight as a function of swelling time.

**Thermogravimetric analysis.** Analysis was performed in Pyris 1 Thermogravimetric Analyzer by PerkinElmer where the content of salt in the hydrogels was validated. Samples of salt solution at 50%$_{sat}$, salt-free hydrogels without initial swelling (as cured), and hydrogels swollen at ambient temperature in the above solutions were placed in an inert pan. The sample size was typically between 10 mg and 40 mg depending on the volume of the sample. All tests were carried out under a current of Argon gas (20 mL min$^{-1}$). The procedure for each sample comprised of three steps: (1) the samples were stabilized for 3 minutes at 30 °C, (2) the samples were heated from 30 °C to 800 °C at a constant rate of 10 K min$^{-1}$, and (3) the samples were held at 800 °C for 5 minutes at the end of the experiment.

**Dynamic vapor sorption experiments.** All the measurements were performed in a DVS RESOLUTION Vapor Gravimetric Sorption Analyzer from Surface Measurements Systems. The method of dynamic vapor sorption (DVS) is applied to determine how much aqueous solution is contained within a solid sample at a constant temperature. Before the measurements, the samples were dried in a vacuum oven at 80 °C for 5 hours. Afterwards, each sample was placed in a chamber at 25 °C under vacuum for 3 hours. During the measurements, the samples are exposed to a controlled relative humidity (RH) atmosphere and the weight is recorded until equilibrium. The time to reach equilibrium may take minutes up to hours for each step. Equilibrium is established when the change in the sample weight did not change by more than 0.005 % min$^{-1}$. During this process, the sample is gravimetrically monitored as a function of the RH. The studied RH ranged from 10% to 90%.

**Cycling experiment.** To test the cyclic hydrogel performance between dry and humid conditions, an oven and a desiccator were used. The desiccator contained a fully saturated NaCl solution at room temperature, which created an environment with a RH of 74%.[34] The oven was set to 60 °C as the dry condition of an RH of practically 0%. The hydrogel was alternately placed in the oven to dry for 8 hours and then in the desiccator to absorb vapor for 16 hours, over the course of two weeks. The sample weight was monitored between conditions.

**Mechanical testing.** Flat punch indentation measurements were performed with a commercial TA.XTplus100C texture analysis instrument from Stable Micro Systems, equipped with a force capacity of 500 g and a 6 mm diameter flat punch indenter. The samples tested were cylindrical discs of polyacrylamide hydrogel, placed on the instrument testing stage. During each measurement, the sample was loaded to a force of 5 grams and then unloaded and the force and displacement of the indenter were recorded. The files with the data were plotted and analyzed with MATLAB, where a slope of the linear range from the beginning of the unloading curve was used to calculate the elastic modulus, based on the Oliver and Pharr theory.[35]

**Scanning electron microscopy characterization.** SEM was conducted with Phenom Pharos Desktop SEM from Phenom World using an acceleration voltage of 10kV and a secondary electron detector. All samples were fully dry and sputtered with a 10 nm coating of gold prior to imaging.

**Acknowledgements**

The authors thank Fonds der chemischen Industrie im Verband der chemischen Industrie e. V. and the Vice President of Research at Humboldt-Universität zu Berlin for financial support. L.H. acknowledges the DAAD RISE Germany program and Northwestern University for financial support. L.H. and M.H. acknowledge the Humboldt-Internship Program for financial support. We thank Keven Walter for the help with the TGA measurements, Carolin Schröter and Lukas Bangert for the help with the mechanical testing measurements, and Carlos D. Díaz-Marín for fruitful discussions.